\documentclass{emulateapj}

\usepackage{graphicx}
\usepackage{amssymb}
\usepackage{amsmath}
\usepackage{epstopdf}
\usepackage{color}
\usepackage{hyperref}

\def \deg {\ensuremath{^{\circ}}}

\def \sgr {SGR\,1806$-$20}
\def \psr {PSR\,J1557$-$4258}

\def \swift {\emph{Swift}}

\def \fermi {\emph{Fermi}}

\shorttitle{Radio Non-Detection of Giant Flare from SGR\,1806$-$20}
\shortauthors{Tendulkar, S.~P. et al.}

\begin{document}

\setlength{\fboxsep}{0pt}%
\setlength{\fboxrule}{0.5pt}%

\title{Radio Non-Detection of the SGR\,1806$-$20 Giant Flare and Implications for Fast Radio Bursts}

\author{Shriharsh P. Tendulkar\altaffilmark{a}, Victoria M. Kaspi\altaffilmark{a}, and Chitrang Patel\altaffilmark{a}}
\email{shriharsh@physics.mcgill.ca}

\altaffiltext{A}{Department of Physics \& McGill Space Institute,\\ 3600 University St,\\ Montr\'eal, QC H3A 2A8, Canada}

\keywords{stars: individual (SGR\,1806$-$20) -- stars: neutron}

\begin{abstract}
We analyse archival data from the Parkes radio telescope which was observing a location 35.6$^\circ$ away from \sgr\ during its giant  $\gamma$-ray flare of 2004 December 27.  We  show that no FRB-like burst counterpart was detected, and set a radio limit of 110\,MJy at 1.4 GHz, including the estimated 70\,dB suppression of the signal due to  its location in the far side lobe of Parkes and the predicted scattering from the interstellar medium. The upper limit for the magnetar giant flare radio to $\gamma$-ray fluence ratio is $\eta_\mathrm{SGR} \lesssim 10^{7}\,\mathrm{Jy\,ms\,erg^{-1}\,cm^{2}}$. Based on the non-detection of a short and prompt $\gamma$-ray counterpart of fifteen FRBs in $\gamma$-ray transient monitors, we set a lower limit on the fluence ratios of FRBs to be $\eta_\mathrm{FRB} \gtrsim 10^{7-9}\,\mathrm{Jy\,ms\,erg^{-1}\,cm^{2}}$. The fluence ratio limit for \sgr\ is inconsistent with all but one of the fifteen FRBs. We discuss possible variations in the magnetar-FRB emission mechanism and observational caveats that may reconcile the theory with observations.    
\end{abstract}

\section{Introduction}
Fast Radio Bursts (FRBs) are bright (0.1--30\,Jy peak flux), millisecond-timescale bursts with an event-rate of $\sim4.4\times10^{3}\,\mathrm{sky^{-1}\,day^{-1}}$ \citep{rane2015} with a minimum fluence of 4\,Jy-ms at 1.4\,GHz.  To date, 16 events have been reported: 14 from the Parkes observatory \citep{lorimer2007, keane2011, thornton2013, burkespolaor2014, petroff2015a, champion2015, ravi2015}, one from the Arecibo radio telescope \citep{spitler2014} and one from the Green Bank telescope \citep{masui2015}. Unlike `perytons' \citep{burkespolaor2011a} which are now known to be local sources of interference \citep{petroff2015b}, FRBs very precisely obey the $\nu^{-2}$ time-delay induced by cold plasma. The hallmark is their dispersion measure (DM), which is much greater (by factors of 3--10) than the DM contribution expected from the Galactic interstellar medium. The excess DM may be intrinsic to the source, placing it within the Galaxy, although this possibility now seems unlikely; it may arise mostly from the intergalactic medium, placing a source of FRBs at cosmological distances ($z\sim0.2-1$) or it may arise from the host galaxy, placing a source of FRBs at extragalactic, but not necessarily cosmological, distances ($\sim100$\,Mpc).

%


Due to FRBs' mysterious nature and the lack of information about them, a plethora of source models have been proposed including Crab-like giant pulses from young extra-galactic pulsars \citep{cordes2015}, planets in pulsar magnetospheres \citep{mottez2014}, neutron-star mergers and the `blitzar' model \citep{totani2013, ravi2014}, black hole-neutron star mergers \citep[][ for a sub-population of FRBs]{mingarelli2015}, magnetar pulse-wind interactions \citep{lyubarsky2014}, flares from nearby stars \citep{loeb2014}, quark novae \citep{shand2015}, and perhaps most popularly, magnetar giant flares \citep{popov2010, popov2013, kulkarni2014, katz2014, katz2015, pen2015, kulkarni2015}. 


\subsection{Magnetar Giant Flares as FRBs}
Galactic magnetars --- neutron stars with strong inferred surface magnetic field strengths ($B_\mathrm{surf}\sim10^{15}$\,G) --- have been observed to emit extremely luminous X-ray and $\gamma$-ray bursts known as giant flares. Since the advent of X-ray and $\gamma$-ray astronomy, we have observed three giant flares: one from SGR\,0526$-$66 on 1979 March 5 \citep{mazets1979}, one from SGR 1900+14 on 1998 August 27 \citep{hurley1999} and one from \sgr\ on 2004 December 27 \citep{palmer2005,hurley2005,terasawa2005}. A giant flare consists of a luminous, sharp, hard X-ray peak, with rise time of order milliseconds, lasting 50--100\,ms, followed by an oscillating tail that is $\sim10^3$ times fainter. 

 Coherent radio counterparts of X-ray bursts from magnetars have been proposed with fluxes as high as 1\,kJy \citep{lyutikov2002,lyutikov2006}. Not long after the announcement of the `Lorimer' burst \citep{lorimer2007}, \citet{popov2010} suggested that such a burst may be a radio counterpart of an extra-galactic magnetar giant flare. With the discovery of more FRBs, multiple authors furthered the explanation, pointing out the good correspondance between the FRB and giant flare energetics, rates and between the observed high DMs and the dense gas in star forming regions \citep{popov2013, lyubarsky2014, katz2015, pen2015, kulkarni2015}. \citet{katz2014} suggested that an FRB from a Galactic magnetar giant flare should be $\mathcal{O}(10^{11})$ times brighter ($\sim$10--100\,MJy) and would be easily discoverable even in the sidelobes of telescopes.

Here we report on archival Parkes radio telescope data obtained coincidentally with the 2004 December 27 giant flare of \sgr.  In Section~\ref{sec:parkes} we use these data to place stringent limits on any radio emission produced by this event, and  in Section~\ref{sec:gamma-ray} we calculate the radio to gamma-ray fluence ratio  with those observed for the 15 FRBs for which this information is available.  As we discuss in Section~\ref{sec:discussion}, our results call into question models in which FRBs arise from magnetar giant flares.

\section{Radio Non-Detection of \sgr\ Giant Flare}
\label{sec:parkes}
The giant flare from magnetar \sgr\ was the brightest $\gamma$-ray event in astronomical history and saturated all the $\gamma$-ray observatories \citep{hurley2005,palmer2005,mazets2005,terasawa2005}. The $\gamma$-ray fluence of the giant flare (in the sharp peak) was estimated to be $\sim2\,\mathrm{erg\,cm^{-2}}$ above 50\,keV \citep{terasawa2005} from the \emph{Geotail} satellite. \citet{hurley2005} estimated the fluence to be $\sim1.4\,\mathrm{erg\,cm^{-2}}$ above $30$\,keV. \citet{palmer2005} estimated the fluence to be $0.8\,\mathrm{erg\,cm^{-2}}$ between 45\,keV and 10\,MeV. In this work, we use a value of $\sim1.4\,\mathrm{erg\,cm^{-2}}$ since it is closer in energy band to the detections of the current $\gamma$-ray instruments, but our conclusion is not sensitive to this range of values.

\subsection{Flare Arrival Time}
The peak of the giant flare crossed the Earth's center at 21:30:25.64 UT \citep{mazets2005}. At the epoch of the $\gamma$-ray flare, \sgr\ was at an altitude of 31.5\deg\ and an azimuth of 95.1\deg\ (East of North) at the Parkes Observatory (longitude = 148.2635101\deg, latitude = $-$32.9984064\deg, altitude=414.80\,m)\footnote{\url{http://www.narrabri.atnf.csiro.au/observing/users_guide/html/chunked/apg.html}}. Based on the geometry, the flare wavefront should have arrived at Parkes 11\,ms before it crossed the center of the Earth, i.e. at 21:30:25.53 UT. These transformations were calculated using the \texttt{astropy.coordinates} and \texttt{astropy.time} routines \citep{astropy2013}.


The telescope started observing pulsar \psr\ at 21:29:19 UT  with the central beam of the Parkes multi-beam receiver and the SCAMP observing system. The telescope was pointed at altitude of 63.1\deg\ and an azimuth of 121.1\deg, 35.6\deg\ away from the location of \sgr. \sgr\ was not hidden behind the telescope feed legs. If the giant flare was accompanied by a prompt radio flare similar to FRBs, we would expect it to arrive 66.5\,s+$t_\mathrm{DM}$ from the beginning of the observation, where $t_\mathrm{DM}$ is the time delay between radio and $\gamma$-ray pulses due to dispersion. 


Based on the distance to \sgr\ \citep[$8.7^{+1.8}_{-1.5}\,$kpc][]{bibby2008} and the NE2001 model \citep{cordes2002}, \citet{lazarus2012} estimated the DM to \sgr\ to be $\sim750\,\mathrm{pc\,cm^{-3}}$. The 90\% upper limit for the distance of \sgr\ was estimated to be 18.6\,kpc \citep{svirski2011} with a corresponding model-predicted DM of 1423$\,\mathrm{pc\,cm^{-3}}$. The model-predicted scattering timescales at 1.4\,GHz are 14\,ms and 56\,ms for distances of 8.7\,kpc and 18.6\,kpc, respectively. We discuss the possibility of the NE2001 model underpredicting the scattering in this direction in Section~\ref{sec:scattering}.

We derive another estimate for the DM from the line-of-sight column density  estimated from X-ray observations \citep[$N_\mathrm{H}=9.7\pm0.1\times10^{22}\,\mathrm{cm^{-2}}$;][]{younes2015}. \citet{he2013} measured the correlation between $N_\mathrm{H}$ and DM to be $N_\mathrm{H}/(10^{20}\,\mathrm{cm^{-2}}) = 0.30^{+0.13}_{-0.09}\,\mathrm{DM/(pc\,cm^{-3})}$. The corresponding DM for \sgr\ is $291^{+126}_{-87}\,\mathrm{pc\,cm^{-3}}$, much lower than the value based on the distance estimates and the NE2001 model.

The dispersion delay between the $\gamma$-ray  and any radio prompt emission at 1.374\,GHz is $t_\mathrm{DM} = 4.148808\,\mathrm{ms} \times (\mathrm{DM/pc\,cm^{-3}}) \times (\nu/\mathrm{GHz})^{-2}$, corresponding to 2.20\,s/1000\,$\mathrm{pc\,cm^{-3}}$. 

As described in the next section, we analysed the Parkes data to search for a bright single burst from \sgr.

\subsection{Archival Parkes Data and Analysis}

\begin{figure*}[htbp]
  \center
  \includegraphics[width=\textwidth,clip=true,trim = 0.9cm 2.3cm 2.1cm 9.8cm]{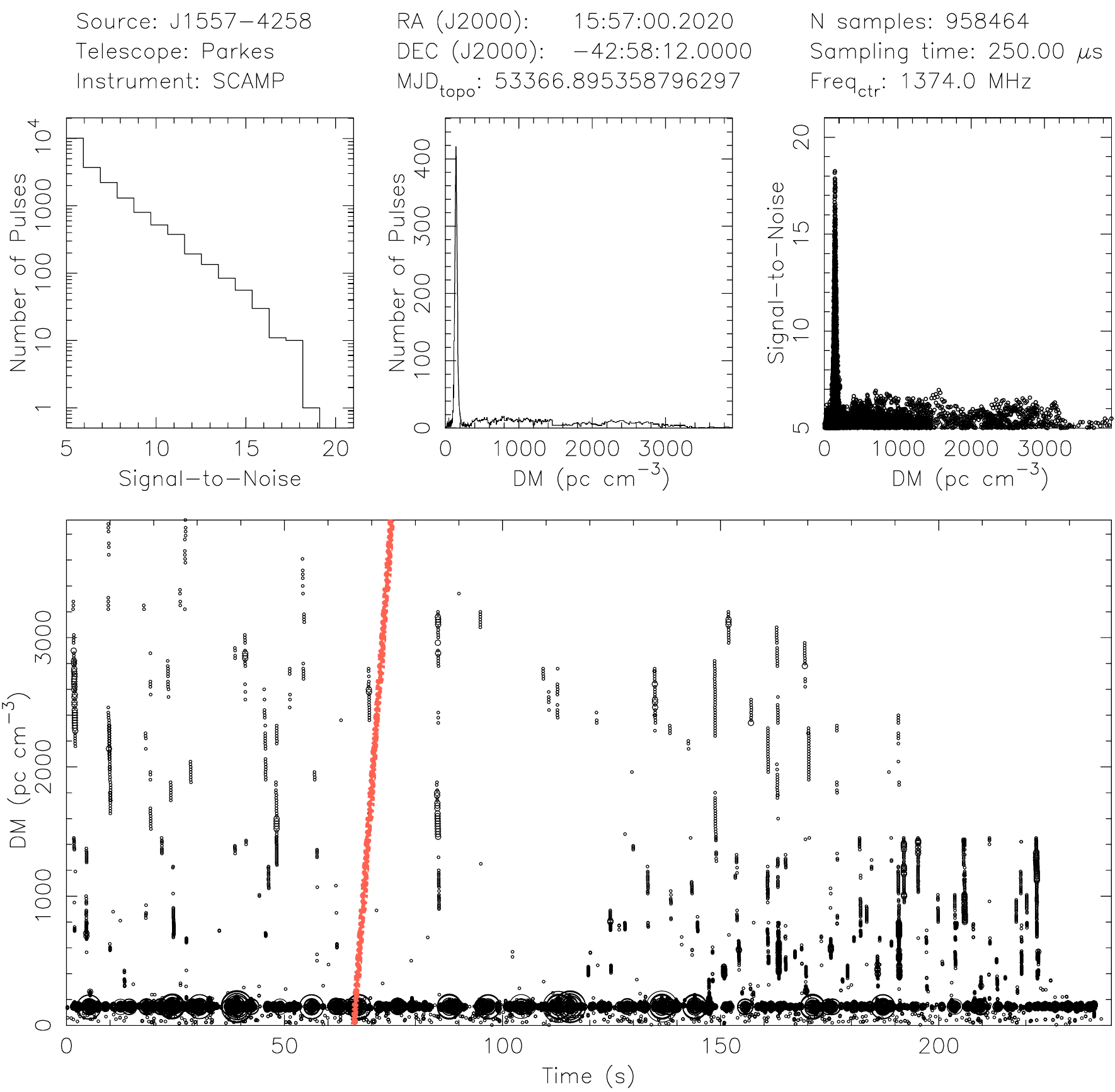}
  \caption{Plots and statistics of dedispersed single pulse events on the 64-m Parkes Telescope data gathered during the \sgr\ giant flare. The bottom plot shows the candidate single pulses detected as a function of time and DM. The symbol size indicates the detection signal to noise ratio (SNR). The flare arrival time corresponds to 66.5\,s after the start of the observation (red track). The slope of the red track corresponds to the expected dispersion delay between the $\gamma$-ray arrival and the pulse arrival at 1374\,MHz. The cluster of candidates to the left of the red track at DM$\approx2600\,\mathrm{pc\,cm^{-3}}$ was verified to be narrow band RFI at 1500\,MHz (single channel). The upper three plots show the detection statistics; \emph{Top Left}: Histogram of candidate detections as a function of SNR. \emph{Top Middle}: Histogram of number of candidate detections as a function of DM. The strong peak at DM=144\,$\mathrm{pc\,cm^{-3}}$ corresponds to pulses from \psr. \emph{Top Right}: Scatter plot of SNR vs DM for each candidate detection.}
  \label{fig:parkes_singlepulse}
\end{figure*}

We downloaded the Parkes data (program number: PT242, data file name: PT0242\_037) from the Parkes data archive and converted from the 1-bit SCAMP raw data format to 8-bit filterbank data using \texttt{filterbank} program from \texttt{SIGPROC}\footnote{\url{http://sigproc.sourceforge.net/}}. The filterbank data were processed with a single pulse searching pipeline based on \texttt{PRESTO}\footnote{\url{http://www.cv.nrao.edu/∼sransom/presto/}} utilities. The pipeline has been described in detail by \citet{spitler2014} and we summarize the data analysis here. The raw data were not cleaned for radio frequency interference (RFI) to avoid accidentally excising bursts. Instead, we visually looked at the time-frequency plots of burst candidates to weed out RFI. 

We created a de-dispersion plan for searching DMs between 0--3940\,$\mathrm{pc\,cm^{-3}}$ with DM steps ranging from 0.5\,$\mathrm{pc\,cm^{-3}}$ at the low DM end to 30\,$\mathrm{pc\,cm^{-3}}$ above 3220\,$\mathrm{pc\,cm^{-3}}$. The upper limit was chosen to be approximately three times the largest predicted Galactic DM, consistent but even broader than the above estimates.  Single pulse candidates are identified in each dedispersed time series using a matched filtering algorithm (\texttt{single\_pulse\_search.py}). This algorithm match-filters the time series with a series of boxcar filters to search for single pulse candidates of widths ranging from 1 to 300 time bins (0.25$\,\mathrm{\mu s}$ to 75\,ms). We grouped the candidates with similar DM and arrival time and made time-frequency (`waterfall') plots for verification. 

Figure~\ref{fig:parkes_singlepulse} shows the results of the single pulse searching algorithm. The circles in the bottom plot denote the detected single pulse candidates as a function of arrival time and DM. The radii of the circles represent the signal to noise ratio (SNR) of the detection. The three upper plots show the statistics of the single pulse candidates as explained in the caption. The line of intense detections at DM=144\,$\mathrm{pc\,cm^{-3}}$ are pulses from \psr\ and serve as a sanity check of our pipeline. The red track shows the expected flare arrival time 66.5\,s+$t_\mathrm{DM}$ after the start of the observation. The cluster of candidates at DM$\approx2600\,\mathrm{pc\,cm^{-3}}$ was verified to be narrow-band RFI at 1500\,MHz. At the horizontal scale of the plot, the width of the track corresponds to $\sim1$\,s and it is unlikely for a radio burst to arrive outside that location unless the radio and $\gamma$-ray emission were separated in emission time. 

With a non-detection of a radio flare from \sgr, we now set limits on the radio flux using the system parameters and by estimating the sidelobe suppression of the Parkes telescope. 

\subsection{Radio Flux Limits}
To calculate the lower limit on the flux of a single pulse detection in the Parkes data, we used the methodology of the Parkes multi-beam pulsar survey \citep{manchester2001} which uses the same instrumentation as the data analyzed here. The limiting flux density is given by  $$S_\mathrm{lim} = \frac{\sigma \beta T_\mathrm{sys}}{G\sqrt{BN_p\tau_\mathrm{obs}}},$$ where $\sigma=1.5$ is a loss factor (from the one-bit sampling as well as other effects), $\beta=6$ is detection SNR threshold, $T_\mathrm{sys}=21$\,K is the system temperature, $G=0.735$\,K/Jy is the telescope gain for the central beam, $B=288\,$MHz is the telescope bandwidth, $N_p=2$ is the number of polarizations  and $\tau_\mathrm{obs}$ is the observing time. Thus, for $\tau_\mathrm{obs}=10$\,ms, a reasonable estimate for a intrinsically narrow pulse scattered at an $e$-folding timescale of 14\,ms, we get a 6-$\sigma$ limit of 0.11\,Jy. For 1-ms and 50-ms bursts, the extreme ranges of scattering, the limits are 0.34\,Jy and 0.47\,Jy, respectively. The corresponding radio fluence limits for 10-ms, 1-ms, and 50-ms timescales are 1.1\,Jy-ms, 0.34\,Jy-ms and 2.4\,Jy-ms, respectively. 

\subsubsection{Scattering Timescales}
\label{sec:scattering}
The 1-bit digitizer backend used in the Parkes observations included a high-pass filter of time constant $\sim$0.9 s \citep{manchester2001}.  The Parkes system thus had reduced sensitivity to signals with rise times comparable or longer than this time constant.  However, this should not be a significant issue when considering a putative radio burst from \sgr\ as the scattering measures, noted above, are predicted to be 14\,ms and 56\,ms for a distance of 8.7 kpc and 18.6 kpc, respectively.  Note that the times refer to the 1/$e$ decay time of a one-sided exponential that is convolved with the signal if scattered, i.e. the rise time should be significantly shorter than these quoted times.  Further, we note that the Parkes multi-beam survey discovered PSR\,J1307$-$6318 which has period 5\,s and duty cycle of 50\% \citep{manchester2001}.  Its pulse rise time is $\sim$40--50\,ms, comparable to the maximum likely scattering time for \sgr\ even if it were as distant as 18.6 kpc. Hence, we conclude that the Parkes high-pass filtering is unlikely to have had a deleterious effect on any radio burst from the \sgr, even for scattering times as long as $\sim$56\,ms. 

We note that if the NE2001 model underpredicts the scattering time in this direction by a factor of 20 or more, the signal may be suppressed by the high-pass filter and may be undetectable. However, we find this to be unlikely from observations of other pulsars in that direction. From the ATNF pulsar catalog \citep{manchester2005}, we compiled a list of pulsars located within a 2\deg\ radius of \sgr\ with DM $> 700\,\mathrm{pc\,cm^{-3}}$, their pulse profiles, and pulse widths at half peak-intensity ($W_\mathrm{50}$) at 1.4\,GHz  \citep{morris2002,hobbs2004}. We find seven pulsars with a DM range of $708.1-932.3\,\mathrm{pc\,cm^{-3}}$. The pulse profiles of four pulsars are symmetrical or show very small scattering tails with $W_\mathrm{50}=13-47$\,ms, suggesting a scattering timescale $\lesssim10$\,ms. Three pulsars have visible scattering tails, the longest having $W_\mathrm{50} = 46.3$\,ms at DM $= 867\,\mathrm{pc\,cm^{-3}}$ (PSR\,1809$-$2004). The corresponding NE2001 scattering timescale estimates for these DM are between 15--22\,ms, corresponding to $W_\mathrm{50}=10-15$\,ms for zero intrinsic width pulses, within a factor of three from the measured values.


\subsubsection{Far Sidelobe Response and Suppression}
\label{sec:suppression}

\begin{figure}
  \center
  \includegraphics[width=0.48\textwidth,clip=true,trim = 0cm 0cm 0cm 0cm]{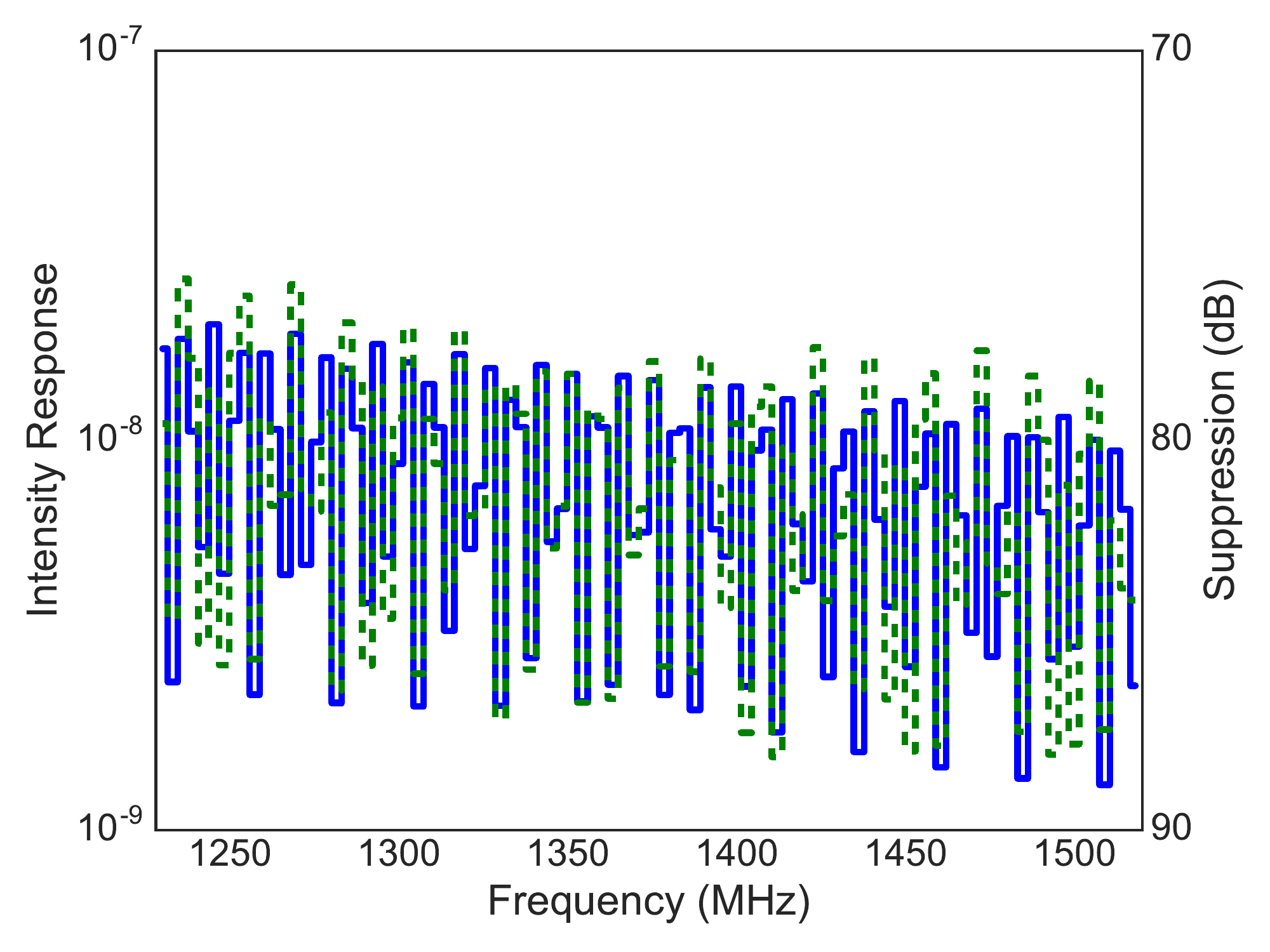}
  \caption{Theoretical sensitivity of an ideal 64-m diameter telescope without and with a 2-m diameter central obstruction (green and blue lines, respectively) as a function of spectral frequency at a position 35.6\deg off-axis. A continuum source will be suppressed by a factor of 80\,dB with respect to the primary beam. The presence of a central obstruction does not significantly change the result.}
  \label{fig:fringes}
\end{figure}

As \sgr\ was 35.6\deg\ away from the pointing center of Parkes, any burst from it would have been detected through the sidelobe response of Parkes, suppressed by many orders of magnitude. Since the far sidelobes are highly suppressed, they are not ordinarily well-characterized. 

The Parkes multi-beam instrument \citep{staveleysmith1996} was primarily designed for 21-cm (1.42\,GHz) HI surveys \citep[for example the Galactic All Sky Survey, GASS; ][]{mccluregriffiths2009} which require careful modelling of the near and far sidelobe response to accurately account for stray radiation \citep{kalberla2010}, albeit at a coarse spatial scale ($\sim$degree). The far sidelobe structure of Parkes is dominated by reflections off the three feed support legs (`stray cones') and radiation from beyond the primary dish (`spillover'). \citet{kalberla2010} modeled the Parkes far sidelobe structure as a constant base level with three stray cones of radius 40\deg\ and width $\pm$3.3\deg. By simultaneous modelling of the sidelobes and deconvolution of the Galactic HI maps, \citet{kalberla2010} fit the suppression of the stray cones and constant base level to be 54\,dB and 70\,dB, respectively, below the primary sensitivity. 

At the time of the flare, \sgr\ was at a location $\theta=35.6\deg,\,\phi=319.9\deg$ in antenna coordinates \citep{kalberla1980}, where $\theta$ is the angular separation from the telescope central beam and $\phi$ is the position angle, measured north through east. This position is 6.4\deg\ (less than one full-width at half maximum) away from a stray cone (P. Kalberla, private communication). The suppression at this position, may be estimated to be between 54\,dB and 70\,dB, $\sim60\pm6$\,dB. We note with caution that these values are based on averaging over $\sim$degree angular scales.

In estimating the strengths of `perytons' which were also detected in the far sidelobes of the Parkes telescope, \citet{burkespolaor2011a} estimate the suppression factor for Parkes sidelobes to be 2500--850000 (corresponding to 34--59\,dB) for the multi-beam receiver. Perytons are known to be near-field due to their coincident detection in all the 13 beams of the receiver while the radio flare from \sgr\ would be in the far-field regime. Even so, the end of the range estimated by \citet{burkespolaor2011a} validates the above estimate. 

We also calculated the diffraction-limited response of an ideal 64-m diameter circular aperture with no obstruction from support legs. At an angular distance of 35.6\deg\ from the center, we calculated the intensity response compared to unity at the central location at each a finely sampled frequency grid over the observing bandwidth. We then binned the intensity measurements over frequency into 3\,MHz channels to simulate the sensitivity of each channel. We also calculated the response including a 2-m diameter central obstruction to account for the receiver block. Figure~\ref{fig:fringes} shows the response of the idealized telescope. In practice, scattering from support legs, reflections from the ground, telescope surface errors will stochastically increase the response in the sidelobes while decreasing the sensitivity in the primary beam. The effect of scattering and reflections is challenging to quantify but it possibly explains the discrepancy between the theoretical suppression compared to the estimate of \citet{burkespolaor2011a}.

Figure~\ref{fig:fringes} also shows that the fringing pattern at large angles varies rapidly in frequency and for a continuum spectrum source, there will be no sharp nulls, i.e. locations in the field of view where the sensitivity drops to zero. Averaging the sidelobe response from Figure~\ref{fig:fringes}, we estimate a sensitivity of $0.98\times10^{-8}$, i.e. a suppression of 80\,dB. The presence of a central obstruction does not significantly change the result. 

We emphasize that the 80\,dB suppression is an idealized upper-limit --- the actual suppression is expected to be between 60--70\,dB.

\subsection{Upper Limit on $\eta_\mathrm{SGR}$}
For a suppression of 60\,dB, 70\,dB, and 80\,dB respectively, the 10-ms radio fluence upper limit of 1.1\,Jy-ms, translates to a fluence limit of 1.1\,MJy-ms, 11\,MJy-ms, and 110\,MJy-ms, respectively. With a $\gamma$-ray fluence of $\sim1.4\,\mathrm{erg\,cm^{-2}}$, we get a 6-$\sigma$ upper limit for $\eta_\mathrm{SGR} < 10^{5.9-6.9}\,\mathrm{Jy\,ms\,erg^{-1}\,cm^{2}}$ for the most likely suppression of 60--70\,dB and $\eta_\mathrm{SGR} < 10^{7.9}\,\mathrm{Jy\,ms\,erg^{-1}\,cm^{2}}$ for a theoretical worst-case suppression of 80\,dB in the idealized diffraction-limited case calculated above. If the scattering timescale or pulse-width ($t$) is larger, the fluence limit scales with $t^{1/2}$.  

\section{Fluence Limit on $\gamma$-ray Bursts from FRBs}
\label{sec:gamma-ray}

\begin{deluxetable*}{lcccclccl}
  \centering
  \tablecolumns{8} 
  \tablecaption{Radio Fluences and $\gamma$-ray Fluence Upper Limits  of FRBs.\label{tab:frb_obs}}
  \tablewidth{0pt}
  \tabletypesize{\footnotesize}
  \tablehead{
    \colhead{Name}  &
    \colhead{Time}  &
    \multicolumn{2}{c}{Coord. (J2000)\tablenotemark{a}} & 
    \colhead{$F_\mathrm{1.4\,GHz}$\tablenotemark{b}}      &
    \colhead{Vis.\tablenotemark{c}}      &
    \colhead{$F_\gamma$\tablenotemark{d}}      &
    \colhead{$\log_{10}(\eta_\mathrm{FRB})$\tablenotemark{e}} &
    \colhead{Ref.}\\
    \colhead{}  &
    \colhead{(UTC)}  &
    \colhead{RA} &
    \colhead{Dec} &
    \colhead{(Jy-ms)}      &
    \colhead{}      &
    \colhead{}      &
    \colhead{} & 
    \colhead{} 
  }
  \startdata 
FRB 010724  & 01-07-24 19:50:00 & 01:18:06 & $   -$75:12:18 & 150.0  & K       & 20& $>8.9$    &\citet{lorimer2007}  \\
FRB 110220  & 11-02-20 01:55:46 & 22:34:38 & $   -$12:23:45 & 8.0    & K       & 20& $>7.6$    &\citet{thornton2013} \\
FRB 130729  & 13-07-29 09:01:49 & 13:41:21 & $   -$05:59:43 & 3.5    & K, B    & 2 & $>8.2$    &\citet{champion2015} \\ 
FRB 010621  & 01-06-21 13:02:09 & 18:52:05 & $   -$08:29:35 & 2.9    & K       & 20& $>7.2$    &\citet{keane2011} \\
FRB 011025  & 01-01-25 00:29:14 & 19:06:53 & $   -$40:37:14 & 2.8    & K       & 20& $>7.1$    &\citet{burkespolaor2014} \\
FRB 131104  & 13-11-04 18:03:59 & 06:44:10 & $   -$51:16:40 & 2.7    & K, G    & 1 & $>8.4$    &\citet{ravi2015} \\ 
FRB 121002  & 12-10-02 13:09:14 & 18:14:47 & $   -$85:11:53 & 2.3    & K, G, B & 1 & $>8.4$    &\citet{champion2015} \\ 
FRB 090625  & 09-06-25 21:53:49 & 03:07:47 & $   -$29:55:36 & 2.2    & K, G, B & 1 & $>8.3$    &\citet{champion2015} \\ 
FRB 110703  & 11-07-03 18:59:38 & 23:30:51 & $   -$02:52:24 & 1.8    & K, G    & 1 & $>8.3$    &\citet{thornton2013} \\
FRB 130626  & 13-06-26 14:55:57 & 16:27:06 & $   -$07:27:48 & 1.5    & K, B    & 2 & $>7.9$    &\citet{champion2015} \\ 
FRB 140514  & 14-05-14 17:14:09 & 22:34:06 & $   -$12:18:46 & 1.3    & K, B    & 2 & $>7.8$    &\citet{petroff2015a} \\ 
FRB 130628  & 13-06-28 03:57:59 & 09:03:02 & $\phd$03:26:16 & 1.2    & K, G, B & 1 & $>8.1$    &\citet{champion2015} \\ 
FRB 121102  & 12-11-02 06:35:53 & 05:32:09 & $\phd$33:05:13 & 1.2    & K       & 20& $>6.8$    &\citet{spitler2014} \\
FRB 110626  & 11-06-26 21:33:15 & 21:03:43 & $   -$44:44:19 & 0.7    & K, G, B & 1 & $>7.8$    &\citet{thornton2013} \\
FRB 120127  & 12-01-27 08:11:20 & 23:15:06 & $   -$18:25:38 & 0.6    & K, B    & 2 & $>7.5$    &\citet{thornton2013}         
 \enddata
\tablenotetext{a}{Pointing location of the telescope at the time of discovery. The FRB positions have an uncertainty of up to a few arc minutes depending on the telescope primary beam size.}
\tablenotetext{b}{Measured radio fluence at 1.4\,GHz. These are lower limits since the FRB may not be detected in the center of the telescope beam.}
\tablenotetext{c}{Visibility to $\gamma$-ray burst instruments. K: Konus-W; G: GBM; B: BAT.}
\tablenotetext{d}{$\gamma$-ray fluence upper limit based on instrument as discussed in the text. Konus-W: $2\times10^{-7}\,\mathrm{erg\,cm^{-2}}$, BAT: $2\times10^{-8}\,\mathrm{erg\,cm^{-2}}$, GBM: $1\times10^{-8}\,\mathrm{erg\,cm^{-2}}$. The best available sensitivity is noted in units of $10^{-8}\,\mathrm{erg\,cm^{-2}}$.}
\tablenotetext{e}{$\eta_\mathrm{FRB} = F_\mathrm{1.4\,GHz}/F_\gamma$ in units of $\mathrm{Jy\,ms\,erg^{-1}\,cm^{2}}$.  }
\end{deluxetable*}

In the papers reporting most FRBs, the authors have searched the available literature or recent GRB Coordinates Network (GCN) messages and reported a non-detection of any counterpart in the X-ray or $\gamma$-ray regime. To set fluence limits on the non-detections, we used the FRB names, radio fluences, epochs and sky locations from the literature (Table~\ref{tab:frb_obs}). For uniformity, we consider only the 15 FRBs that have been detected at ~1.4\,GHz. We have not included FRB\,110523 \citep{masui2015} since it was detected at 800\,MHz and it is not possible to reasonably convert its fluence to 1.4\,GHz given the uncertainty in its intrinsic spectral index.

\subsection{\textit{FERMI} GRB Burst Monitor}
The \textit{FERMI}-GRB Burst Monitor (GBM) instrument consists of 12 Na-I detectors and two Bi-Ge scintillators \citep{meegan2009}. These detectors are sensitive to the entire unocculted sky for photon energies from 8\,keV to 40\,MeV. \textit{FERMI}-GBM data have been available since 2008 February. As \textit{FERMI} is in a low-Earth orbit with an average altitude of 550\,km, the detections of GRBs are limited by Earth occultation. For each FRB, we analysed the visibility of FRB sky location to \textit{FERMI}  using the spacecraft position history files for a duration of 10\,minutes before and after the epoch of the burst to account for any difference in arrival times (Table~\ref{tab:frb_obs}). We find that 6 of the 15 bursts were visible to GBM when they were detected in the radio telescopes.

The on-board trigger threshold of GBM for short bursts is 0.74\,$\mathrm{photons\,cm^{-2}\,s^{-1}}$ \citep{meegan2009AIPC}\footnote{The current threshold is 0.61\,$\mathrm{photons\,cm^{-2}\,s^{-1}}$. \url{http://f64.nsstc.nasa.gov/gbm/instrument/}} corresponding to a fluence of $2\times10^{-9}\,\mathrm{erg\,cm^{-2}}$ for a nominal photon energy of 20\,keV and a burst duration of 100\,ms, characteristics similar to those of magnetar giant flares. As a conservative estimate for detection completeness, we take the fluence limit to be 5 times higher, i.e. $1\times10^{-8}\,\mathrm{erg\,cm^{-2}}$. This value is corroborated by the faintest short GRB fluences listed in the GBM GRB catalog \citep{vonkienlin2014}. 

\subsection{\textit{SWIFT}-BAT}
The Burst Alert Telescope (BAT) instrument aboard the \swift\ satellite (launched 2004 November) is a coded aperture mask high-energy (15--150\,keV) X-ray telescope \citep{barthelmy2005}. The half coded field of view (FoV) is $100^{\circ} \times 60^{\circ}$ (1.4\,sr). Using NASA's HEASARC database, we checked if the BAT was pointing within a radius of $40^{\circ}$ of each FRB between 10 minutes before and after the FRB epoch. We find that 8 of the 15 bursts were within the BAT half coded FoV when they were detected in the radio telescopes.

The design burst flux sensitivity of the BAT is $10^{-8}\,\mathrm{erg\,cm^{-2}\,s^{-1}}$ \citep{barthelmy2005}. For a 100-ms SGR-like burst, this corresponds to a 8-$\sigma$ fluence sensitivity of $8 \times 10^{-9}\,\mathrm{erg\,cm^{-2}}$. The short GRBs detected by \swift-BAT are also used to calculate an upper limit on the fluence. From Table 1 of \citet{sakamoto2011} and Table 1 of \citet{berger2014} we estimate that the short GRBs (timescale $<2$\,s) are detected at a 6-$\sigma$ significance at fluences of $\sim2\times10^{-8}\,\mathrm{erg\,cm^{-2}}$.

\subsection{Konus-\textit{WIND}}
Konus-W is a $\gamma$-ray spectrometer aboard the \textit{GGS-WIND} mission. It has two NaI(Tl) scintillators providing omni-directional sensitivity to $\gamma$-ray bursts between 10\,keV to 10\,MeV \citep{aptekar1995}. The mission was launched in 1994 November and travelled to its final location at the Earth-Sun L1 point (1.5 million km from Earth) in 2004. At this separation, the Earth covers a negligible area (0.2\,sq.\,deg) in the Konus-W sky and hence the Konus scintillators have an essentially unocculted view of the entire sky.

The design sensitivity of Konus-W is 1--5$ \times 10^{-7}\,\mathrm{erg\,cm^{-2}}$ \citep{aptekar1995} at a 6-$\sigma$ level. \citet{mazets2004} lists all the GRBs detected by Konus-W between 1994 and 2002. The faintest GRB detected had a fluence of $\sim2\times10^{-7}\,\mathrm{erg\,cm^{-2}}$. \citet{svinkin2015} calculated the sensitivity of Konus-W to a scaled giant flare from \sgr\ to be $2-5.7\times10^{-7}\,\mathrm{erg\,cm^{-2}}$ (9-$\sigma$) for the known range of spectral parameters. Due to its unocculted view, we assume the Konus-W sensitivity as the fluence lower limit for $\gamma$-ray counterparts for any FRB not in the GBM or BAT FoV.

\subsection{Lower Limits on $\eta_\mathrm{FRB}$}
We define $\eta = F_{\mathrm{1.4\,GHz}}/F_{\gamma}$ as the ratio of burst fluence in the 1.4\,GHz (radio) and $\gamma$-ray bands. Based on the individual $\gamma$-ray fluence limits and the observed radio fluences, we find the $\sim$6-$\sigma$ lower limits on the FRB radio to $\gamma$-ray fluence ratio $\eta_\mathrm{FRB}$ to be between $10^{6.8-8.9}\,\mathrm{Jy\,ms\,erg^{-1}\,cm^{2}}$ (Table~\ref{tab:frb_obs}).

\section{Discussion}
\label{sec:discussion}

\begin{figure*}
  \center
  \includegraphics[width=0.85\textwidth]{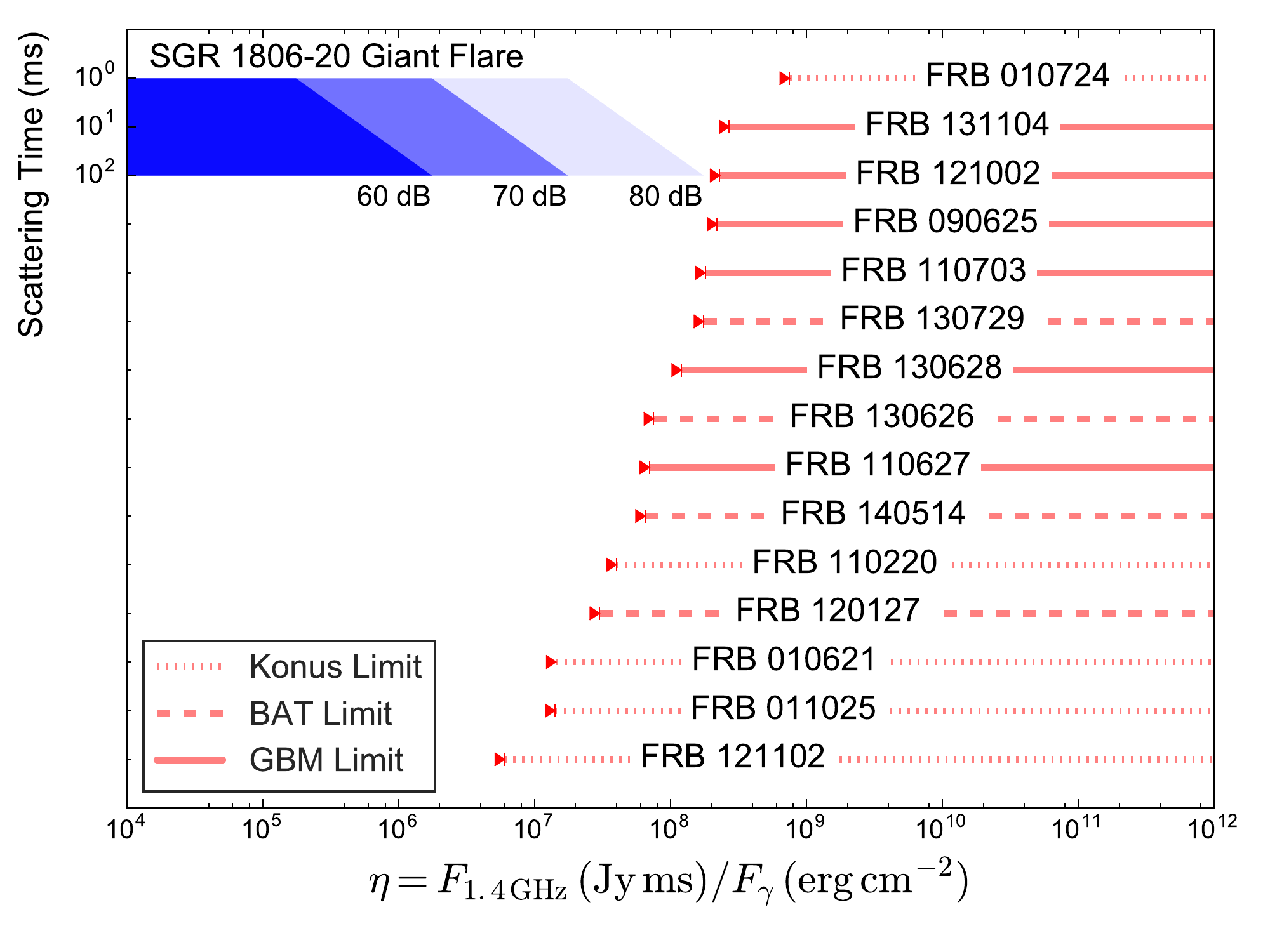}
  \caption{Plot of the ratio of the 1.4\,GHz fluence ($F_\mathrm{1.4\,GHz}$) to the $\gamma$-ray fluence  ($F_\gamma$) from the giant flare of \sgr\ compared to those of fifteen FRBs. The red arrowheads denote the lower limits on $\eta_\mathrm{FRB}$ from the non-detection of $\gamma$-ray counterparts for FRBs based on the sensitivity of \textit{Konus-WIND} (dotted lines), \swift-BAT (dashed lines) and of \textit{FERMI}-GBM (solid lines) from Table~\ref{tab:frb_obs}. The blue trapeziods denote allowed range of $\eta_\mathrm{SGR}$ given the non-detection of a radio flare from \sgr\ for a range of scattering timescales from 1--100\,ms (upper left axis, increasing downwards) for three different sidelobe suppressions. The scattering timescale estimated from NE2001 is $\approx10\,$ms. The 60\,dB, 70\,dB and 80\,dB labels denote possible sidelobe suppressions (see Section~\ref{sec:suppression} for details). The color density signifies the likelihood, with the 60 -- 70\,dB suppression being most likely and the 80\,dB suppression being extremely unlikely. }
  \label{fig:eta_plot}
\end{figure*}

In Section~\ref{sec:parkes}, we have derived an upper limit on the radio to $\gamma$-ray fluence ratio, $\eta_\mathrm{SGR} = F_\mathrm{1.4\,GHz}/F_\gamma$, for the 2004 December 27 giant flare of \sgr\ based on the measured $\gamma$-ray fluence and a radio non-detection in the 64-m Parkes telescope for scattering times estimated from the NE2001 model. In Section~\ref{sec:gamma-ray}, we have placed $\sim$6-$\sigma$ lower limits on $\eta_\mathrm{FRB}$ as summarized in Table~\ref{tab:frb_obs} based on the non-detections of prompt $\gamma$-ray counterparts to the reported FRBs. 

Figure~\ref{fig:eta_plot} summarizes our result, plotting $\eta_\mathrm{FRB}$ and the suppression-dependent values of $\eta_\mathrm{SGR}$ for a range of scattering timescales or intrinsic pulse widths. Assuming the likely suppression for the sidelobe of the Parkes telescope (70 dB) and the scattering timescales from the NE2001 model, all but one of the FRBs have $\eta_\mathrm{FRB}>\eta_\mathrm{SGR}$ by a factor of 1.6--100  with the highest $\eta$ ratio being that of the `Lorimer' burst (FRB\,010724). Even if the scattering timescale is assumed to be 100\,ms and the suppression of the telescope is assumed to be the theoretical worst-case value of 80\,dB, the non-detection of a radio counterpart to the giant flare from \sgr\ is at odds with four of the fifteen known FRBs.

Thus our result challenges the hypothesis that the same event mechanism that produces a magnetar giant flare (at least for \sgr) also produces the known FRBs simultaneously. 

However, in the next section, we explore caveats and reasons why this naive conclusion may not exclude magnetar giant flares as the origins of FRBs. This work does not have an obvious applicability to any of the other explanations that have been suggested, hence we limit our discussion to the magnetar giant flare hypothesis.

\subsection{Variability of Observed $\eta$}

A straightforward way to reconcile the magnetar giant flare hypothesis with the lack of detectable radio emission from the giant flare of \sgr\ is to propose that the observed fluence ratio, $\eta_\mathrm{obs}$, can vary wildly between different magnetars and even different bursts of the same magnetar. This could either be due to the intrinsic variation of the emission mechanism or due to beaming of the radio pulse toward or away from the observer. In this scenario, extragalactic FRBs would be a sub-population of events that are radio bright and/or are beamed towards us. 

The recent discovery of multiple radio bursts from the source of FRB\,121102 \citep[][]{spitler2016,scholz2016} with varied spectral characteristics supports this possibility. The bursts are clustered in time, separated by as little as few tens of seconds, and show diverse spectral slopes and amplitudes. The repetition and the drastic variation in spectral shapes proves that the emission mechanism in this source is definitely not cataclysmic, is able to summon up the required energy within few-minute timescales, and produces emission with varying characteristics. While it is not yet clear whether the source of FRB\,121102 is unique or representative of the FRB population, it is clear that at least in one case the burst properties vary over a wide range. 

In the next section, we discuss a possible caveat on the hypothesis that only a small fraction of giant flares are radio bright based on the current estimates of rates of FRBs and magnetar giant flares. 

\subsubsection{Event Rates of FRBs and Giant Flares}
\label{sec:event_rates}
\begin{figure}
  \center
  \includegraphics[width=0.48\textwidth]{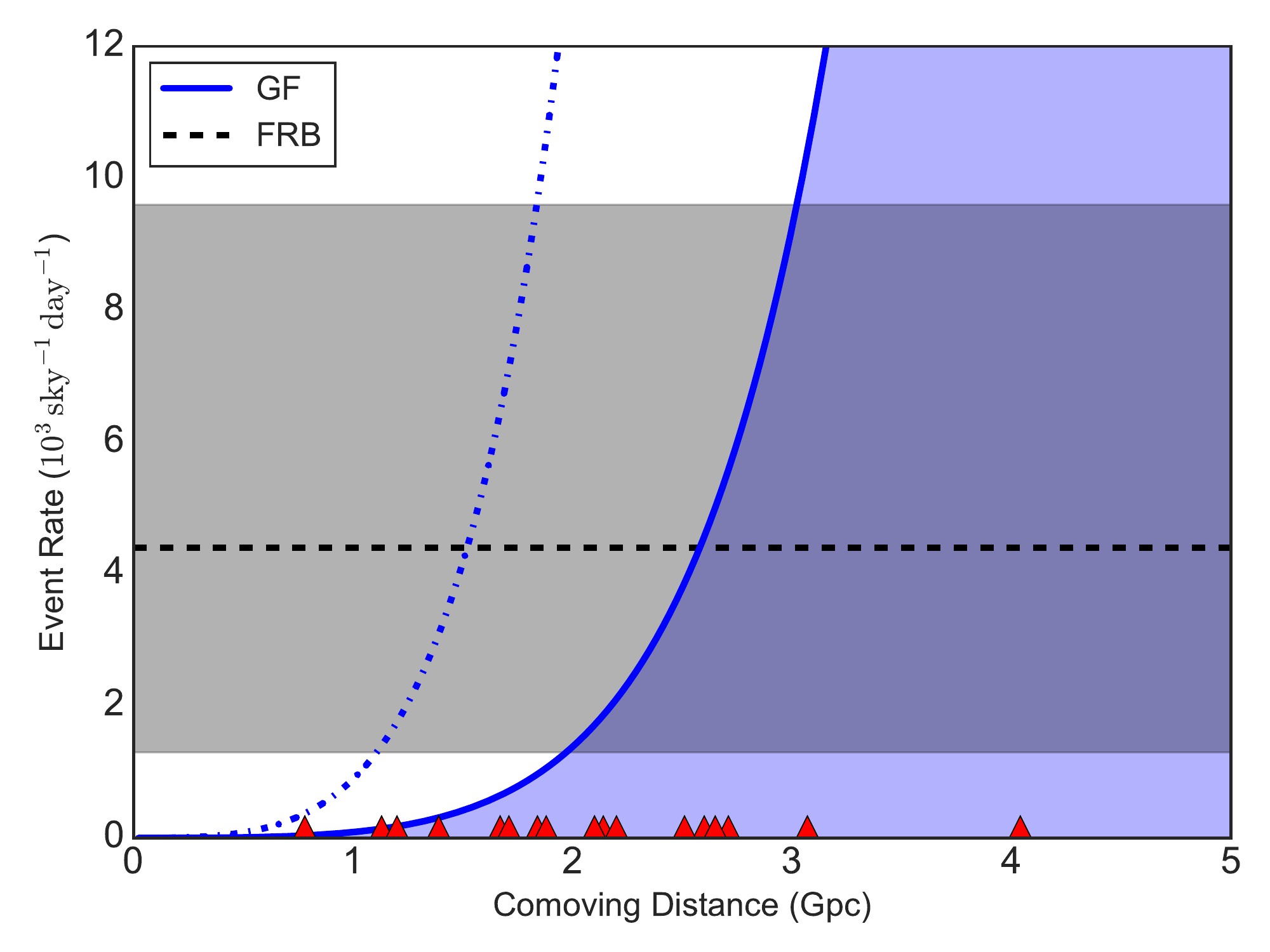}
  \caption{The sky rate of \sgr-like giant flares ($\mathcal{R}_\mathrm{GF}(z)$, solid blue line) as a function of the comoving radius of the sample volume using Equation~\ref{eqn:rate_calc}. The upper limit on the rate of $10^{45}$\,erg giant flares is also plotted (blue dash-dot line). See Section~\ref{sec:event_rates} for a detailed discussion. The dashed black line and the gray shaded region denotes the sky rate of FRBs ($\mathcal{R}_\mathrm{FRB}= 4.4^{+5.2}_{-3.1} \times 10^3\,\mathrm{d^{-1}\,sky^{-1}}$) with a minimum fluence of 4\,Jy-ms at 1.4\,GHz \citep{rane2015}. The red triangles at the bottom denote the upper limits on the distances \citep[from FRBCAT;][]{petroff2016} estimated as $\mathrm{DM_{excess}}=1200 \times z\,\mathrm{pc\,cm^{-3}}$ \citep[see ][]{ioka2003}.}
  \label{fig:frb_rates}
\end{figure}

Let $\epsilon \in [0,1]$ be the fraction of magnetar giant flares that produce observable bright radio bursts with observed $\eta_\mathrm{obs}\gtrsim10^{8-9}$. If $\epsilon\ll 1$, say $\epsilon\lesssim 0.1$, then the lack of a radio counterpart to the giant flare of \sgr\ is not surprising. However, if $\epsilon \approx1$, then the absence of a radio burst at the time of the giant flare from \sgr\ is significant. By definition, $\epsilon$ accounts for intrinsic variation in $\eta$ as well as the beaming towards the Earth. Therefore, we can write $$\mathcal{R}_\mathrm{FRB}(z) = \epsilon \mathcal{R}_\mathrm{GF}(z),$$ where $\mathcal{R}_\mathrm{FRB}(z)$ is the all-sky rate of FRBs observed from sources up to a redshift $z$, corresponding to a comoving radial distance $d(z)$, and $\mathcal{R}_\mathrm{GF}(z)$ is the rate of SGR giant flares in the same volume.

Due to small number statistics, the uncertainties in the estimated rates of FRBs and \sgr-like giant flares are quite large. In this section, we make an effort to understand and incorporate all the uncertainties that are within the scope of this paper.

The rate of giant flares can be estimated by 
\begin{equation}
\label{eqn:rate_calc}
\mathcal{R}_\mathrm{GF}(z) = \int_{z^\prime=0}^{z^\prime=z}\mathrm{d}V(z^\prime)\times\Gamma_\mathrm{CCSN}(z^\prime)\times f_\mathrm{M}\times \tau_\mathrm{active} \times r_\mathrm{GF},
\end{equation}
 where $\mathrm{d}V(z^\prime)$ is the differential comoving volume element at a redshift $z^\prime$, $\Gamma_\mathrm{CCSN}(z^\prime)$ is the volume rate of core-collapse supernovae as a function of redshift, $f_\mathrm{M} \approx 0.1$ is the fraction of the magnetar birthrate as compared to the pulsar birthrate \citep{keane2008}, $\tau_\mathrm{active} \approx 5\,\mathrm{kyr}$ is the active lifetime of the magnetar for emitting giant flares and $r_\mathrm{GF}$ is the rate of \sgr-like giant flares per magnetar.

The volume rate of core-collapse supernovae increases with increasing redshift as $ \Gamma_\mathrm{CCSN}(z^\prime) = \Gamma_0 (1 + z^\prime)^\beta$, $\beta\approx4.3$ due to an increasing star-formation rate \citep[see ][]{taylor2014, karim2011}. The estimated volume rate at $z=0$ is $\Gamma_0\approx 0.71\pm0.15\times 10^{-4}\,\mathrm{yr^{-1}\,Mpc^{-3}}$ \citep{li2011}. 

The rate of giant flares per magnetar, $r_\mathrm{GF}$, is highly uncertain. As mentioned before, only three giant flares have been observed since the launch of the Vela satellites in mid-1960's. The giant flares from SGR\,0526$-$66 and from SGR\,1900+14 had total energy outputs of $\lesssim10^{44-45}\,$erg \citep{fenimore1996,feroci2001} while the giant flare from \sgr\ was an order of magnitude more luminous \citep[$\approx2\times10^{46}$\,erg][]{palmer2005}. We know of 23 magnetars in the Milky Way and the Large Magellanic Cloud \citep{olausen2014} and we have been sensitive to giant flares for about 50 years. Using the 95-percent confidence limits from \citet{gehrels1986}, we get $r_\mathrm{GF,46}=(0.02-5)\times10^{-3}\,\mathrm{magnetar^{-1}\,yr^{-1}}$ for \sgr-like (energy $\approx10^{46}$\,erg) giant flares or $r_\mathrm{GF,45}=(0.5-8)\times10^{-3}\,\mathrm{magnetar^{-1}\,yr^{-1}}$ for giant flares with energies $\approx10^{44-45}\,$erg. 

 $r_\mathrm{GF}$ has also been constrained from the rates of nearby short GRBs that are consistent with being giant flares from extragalactic magnetars \citep{ofek2007,svinkin2015}. The \textit{Konus}-WIND experiment is sensitive to \sgr-like giant flares up to a distance of $\approx30\,$Mpc and to giant flares of $\approx10^{45}\,$erg  up to a distance of $\approx6\,$Mpc. From a sample of 147 short GRBs detected by the \textit{Konus}-Wind experiment, \citet{svinkin2015} calculated a one-sided 95-percent upper limit of $r_\mathrm{GF,46} \leq (0.6-1.2) \times 10^{-4}\,\mathrm{yr^{-1}\,magnetar^{-1}}$ for giant flares as powerful as the one from \sgr. The range of values reflects the uncertainty in our understanding of the star formation rate in our Galaxy and in the local Universe. For less energetic flares with energy output $\lesssim10^{45}$\,erg, the upper limit on the rate was calculated to be $r_\mathrm{GF,45} \leq (0.9-1.7)\times 10^{-3}\,\mathrm{yr^{-1}\,magnetar^{-1}}$. These upper limits are consistent with and more constraining than the rates calculated above from the Galactic giant flares.

Figure~\ref{fig:frb_rates} compares the estimated rate of FRBs with the estimated rates of giant flares occuring within a given comoving radius, calculated by integrating Equation~\ref{eqn:rate_calc}. The dashed black line and the gray shaded area show the current estimate of the sky rate of FRBs with a minimum fluence of 4 Jy-ms at 1.4\,GHz, $\mathcal{R}_\mathrm{FRB}= 4.4^{+5.2}_{-3.1} \times 10^3\,\mathrm{d^{-1}\,sky^{-1}}$\citep{rane2015}. The solid blue line and the shaded blue region represent the rates of \sgr-like giant flares with $r_\mathrm{GF,46}\lesssim 1 \times 10^{-4}\,\mathrm{yr^{-1}\,magnetar^{-1}}$. The dash-dot blue line represents the upper limit on the rate of $\approx10^{45}$\,erg giant flares, $r_\mathrm{GF,45}\lesssim 1 \times 10^{-3}\,\mathrm{yr^{-1}\,magnetar^{-1}}$.  

The distances to the known FRBs can be estimated from their DM if we can separate the contribution of the ionized inter-galactic medium ($\mathrm{DM_{IGM}}$). Since the contribution of the FRB host galaxy is not known, an upper limit for $\mathrm{DM_{IGM}}$ is $\mathrm{DM_{excess} = DM_{FRB} - DM_{MW}}$, where $\mathrm{DM_{MW}}$ is the best estimate of the Milky Way contribution along the line of sight. The FRB Catalog \citep[FRBCAT][]{petroff2016} lists the $\mathrm{DM_{excess}}$ and the corresponding distance upper limits for the known FRBs assuming a DM to redshift conversion of $\mathrm{DM_{IGM}}=1200z\,\mathrm{pc\,cm^{-3}}$ \citep[see ][]{ioka2003}. The highest $\mathrm{DM_{excess}}=1555\,\mathrm{pc\,cm^{-3}}$ is for FRB\,121002 corresponding to a comoving distance upper limit of 4\,Gpc, while most known FRBs have a $\mathrm{DM_{excess}}$ in the range 500--900\,$\mathrm{pc\,cm^{-3}}$, corresponding to distance upper limits of 1.5--3\,Gpc. In Figure~\ref{fig:frb_rates}, the red triangles denote the upper limits on the distances to known FRBs calculated from $\mathrm{DM_{excess}}$. 

Note that FRB\,121002 was detected with a signal to noise ratio of 16 with a fluence of 2.3\,Jy-ms, rather than as a marginal detection. This suggests that the DM distance scale to FRBs is not currently limited by our sensitivity. If that was the case, we would expect more of the detected FRBs to be from further away as they probe a larger volume and because the star formation rate increases rapidly with redshift. Thus, the distance scale to currently observed FRBs is probably not significantly larger than $\sim$ 3\,Gpc and possibly lower when DM contributions of the hosts are included. 

If the distance scale of FRBs is $\sim2-3$\,Gpc and FRBs are related to \sgr-like giant flares from magnetars, then it is clear from Figure~\ref{fig:frb_rates} that the rates of FRBs and giant flares are likely comparable, i.e. it is unlikely that $\epsilon \ll 1$. This suggests that the non-detection of a radio counterpart of the giant flare from \sgr\ is significant. However, if we consider the rates of weaker giant flares ($10^{44-45}$\,erg) and the more frequent intermediate flares \citep[for a review, see ][]{turolla2015} at energy scales of $10^{38-40}$\,erg, then the observed rate of FRBs may be consistent with being a small fraction of the magnetar activity rate. We also note that the distance estimates to FRBs shown in Figure~\ref{fig:frb_rates} are upper limits and neglect any DM contribution from the FRB host galaxy, hence the true $\epsilon$ may be higher. 

\subsubsection{Circum-magnetar Environment}
\citet{lyubarsky2014} proposed a model for FRBs arising from the interaction of a magnetic shock with the wind nebula of a magnetar. In this model, the giant flare causes a strong magnetic perturbation to propagate outward into the nebula created by the pulsar wind in the circum-magnetar material. As the perturbation interacts with the density discontinuity at the edge of the nebula, the particles accelerated by the shocks can produce a burst of broadband synchrotron maser emission which is observed as an FRB. The radio efficiency of such a process is expected to be $10^{-5}-10^{-6}$, which can explain observed FRB energies ($10^{38-40}\,$erg) from magnetar giant flares.

If the magnetar lacks a significant wind nebula and the corresponding density discontinuity, this mechanism may not operate, creating a $\gamma$-ray flare without a corresponding FRB, which may be the case for the giant flare from \sgr. However, the rate calculation presented above can also be applied in this case --- if the fraction of magnetar giant flares that lead to FRBs is small, then the FRB distance scale may need to be larger, the rate of magnetar flares may need to be higher or the rates of FRBs may need to be lower. We also note that our search was only for radio and $\gamma$-ray bursts that are simultaneous (after correcting for the dispersion delay) within a window of $\sim$ minute timescales. It is possible in this model for the radio flare to be delayed from the  $\gamma$-ray flare due to the difference in the velocity of light (for the $\gamma$-ray flare propagation) and the Alfv\'en velocity (for the propagation for the magnetic disturbance) through the wind nebula, depending on the plasma density and the magnetic field strength.



\subsection{Filtering and Mis-identification of Galactic FRBs?}
If FRBs arise from magnetars and the rates of FRBs are comparable to the rates of giant flares, it is worth asking if FRBs from within the Milky Way could have been observed  through the history of radio astronomy. A Galactic FRB, detected far off-axis in multi-feed telescopes such as Parkes and Arecibo, will likely have coincident detections in all feeds since the far sidelobe sensitivity of all beams is similar. Pulsar and transient surveys utilize the the multi-feed time streams to mask out near-field radio frequency interference (RFI), including perytons. The real-time FRB pipeline of the High Time Resolution Universe (HTRU) survey at Parkes uses two FRB detection criteria that would rule out the detection of Galactic FRBs: (a) $\mathrm{DM \geq 1.5\times DM_{MW}}$ and (b) $N_\mathrm{beams} \leq 4$, where $\mathrm{DM_{MW}}$ is the Galactic DM contribution along the line of sight and $N_\mathrm{beams}$ is the number of beams (out of a total of 13) in which the burst is detected \citep{petroff2015a}. The searches for Galactic bursts used an $N_\mathrm{beams} \leq 9$ criteria \citep{burkespolaor2011b}. The PALFA survey strategy is to subtract the feed-averaged zero DM time series for transient searches (Patel, C. et al, in preparation). This would significantly reduce the sensitivity of PALFA to low-DM Galactic FRBs.

Unlike `perytons' however, Galactic FRBs would have a sharp $\nu^{-2}$ dispersion sweep, but the DM would be within the Galactic limit. A single-beam radio telescope, without anti-coincidence filtering, may misidentify these bursts as candidate rotating radio transients \citep[RRATs; ][]{keane2015} in the pointing direction. A revised search of archival multi-feed data optimized for detecting Galactic FRBs with coincident detections in almost all beams and a lower DM thresholds may set limits on the rate of occurence. Another feature of these sidelobe mis-identifications is that radio pulses classified as unconfirmed RRATs at different sky locations may appear to be clustered in DM space since they could originate from the same Galactic magnetar. A preliminary scan through the RRATalog\footnote{\url{http://astro.phys.wvu.edu/rratalog/}} for RRATs detected only in single observations shows no obvious groupings, though a firm statistical analysis is complicated due to the heterogenous surveys that have contributed to the detections and is beyond the scope of this paper.

It is worth mentioning that Galactic FRBs, especially due to their enormous flux and small dispersion, may be detectable not only by astronomical telescopes but also by civilian and military radars, atmospheric monitoring experiments, ground-satellite communication links etc. That such a phenomenon has not yet been reported suggests to us it is unlikely to exist.

\subsection{Magnetar Ages and Behavior}
It is also possible that \sgr\ is a special case and that the `typical' magnetars that produce FRBs have different characteristics. Indeed, \sgr\ is the youngest magnetar known, with a characteristic age\footnote{From the McGill Magnetar Catalog: \url{http://www.physics.mcgill.ca/~pulsar/magnetar/main.html}}  of 240\,yr \citep{olausen2014} and a kinematic age of 650\,yr \citep{tendulkar2012}, has the strongest $B$-field of all those known \citep[$B_\mathrm{surf}\approx2\times10^{15}$\,G; ][]{kouveliotou1998,woods2007}, and the giant flare of 2004 December 27 was the most luminous of all three observed giant flares\footnote{While the SGR\,1900+14 giant flare of 1998 August 27, was visible from Parkes, the telescope was undertaking maser observations at 22\,GHz, rendering it unusable for burst detections.}.

It is possible that a very young ($<<100$\,yr, `baby') magnetar may be extremely active in emitting energetic  radio-bright, $\gamma$-ray faint bursts. As the magnetar ages, the repetition may slow and $\eta$ may decrease to a point where the bursts are infrequent ($<$ once per decade) and radio faint. While this may explain the lack of a Galactic FRB, however, it also significantly reduces the number of extragalactic magnetars that are available to explain the FRB rate.

Another possibility is that $\eta$ may be suppressed significantly by a strong $B$-field, implying an upper bound for the $B$-field of FRB sources. In such a case, we may expect the other, lower $B$-field, Galactic magnetars to produce Galactic FRBs, which again, would be detectable in sidelobes of telescopes.

\section{Conclusion}
 From our non-detection of a prompt radio counterpart to the 2004 December 24 giant flare of \sgr\ for predicted scattering times, we have calculated an upper limit on the radio to $\gamma$-ray fluence ratio, $\eta_\mathrm{SGR}$ and shown that it is inconsistent with $\eta_{FRB}$ for fourteen of the fifteen FRB sources for which a lower limit on $\eta$ is calculated from  their non-detection in \fermi-GBM, \swift-BAT and Konus-\emph{WIND} $\gamma$-ray instruments. The result challenges the simple hypothesis that magnetar giant flares and FRBs are the prompt multi-wavelength counterparts of the same emission mechanism. We have discussed the implications of the result and the possible modifications to this hypothesis that may reconcile the two phenomena which are very similar in other aspects: rates, energetics, and timescales. 

\acknowledgements
We thank the anonymous referee, J. Katz, K. Postnov and M. Bailes for helpful comments and discussions. We would like to thank Lister Staveley-Smith and Peter Kalberla for helpful discussions regarding the Parkes telescope and Dick Manchester, Lawrence Toomey, and Vincent McIntyre for help in accessing archival Parkes data. This research made use of Astropy, a community-developed core Python package for Astronomy \citep{astropy2013}.

S.P.T acknowledges support from a McGill Astrophysics postdoctoral fellowship. V.M.K. acknowledges support from an NSERC Discovery Grant and Accelerator Supplement, funds from the Centre de Recherche Astrophysique du Quebec, the Canadian Institute for Advanced Research, a Canada Research Chair and the Lorne Trottier Chair in Astrophysics \& Cosmology.

\bibliographystyle{apj}
\bibliography{paper}

\end{document}